\documentclass[journal]{IEEEtran}
\IEEEoverridecommandlockouts
\usepackage{cite}
\usepackage{amsmath,amssymb,amsfonts}
\usepackage{bbm}
\usepackage{algorithm,algorithmic}
\usepackage{graphicx}
\usepackage{epstopdf}
\usepackage{textcomp}
\usepackage{xcolor}
\def\BibTeX{{\rm B\kern-.05em{\sc i\kern-.025em b}\kern-.08em
		T\kern-.1667em\lower.7ex\hbox{E}\kern-.125emX}}
\usepackage{float}
\usepackage{amsmath}
\usepackage{caption}
\newtheorem{pb}{Problem}
\newtheorem{theorem}{Theorem}
\newtheorem{lemma}{Lemma}
\newtheorem{definition}{Definition}

\newtheorem{corollary}{Corollary}

\graphicspath{{figures/}}
\allowdisplaybreaks
\begin{document}
	
	\title{Scheduling to Minimize Age of Information in Multi-State Time-Varying Networks with Power Constraints}
	\author{
		\IEEEauthorblockN{Haoyue~Tang\textsuperscript{1},~Jintao~Wang\textsuperscript{1,2},~Linqi~Song\textsuperscript{3},~Jian~Song\textsuperscript{1,2}}
		
		\IEEEauthorblockA{
			\textsuperscript{1}Beijing National Research Center for Information Science and Technology (BNRist),\\
			Dept. of Electronic Engineering, Tsinghua University, Beijing 100084, China\\
			\textsuperscript{2}Research Institute of Tsinghua University in Shenzhen, Shenzhen, 518057\\
			\textsuperscript{3}Department of Computer Science, City University of Hong Kong, Kowloon Tong, Hong Kong\\
			\{thy17@mails, wangjintao@,jsong@\}tsinghua.edu.cn, linqi.song@cityu.edu.hk}}
	
	\maketitle
	
	\begin{abstract}
		In this paper, we study how to collect fresh data in time-varying networks with power constrained users. We measure data freshness from the perspective of the central controller by using the metric \emph{Age of Information}, namely the time elapsed since the generation time-stamp of the freshest information. We wonder what is the minimum AoI performance the network can achieve and how to design scheduling algorithms to approach it. To answer these questions when scheduling decisions are restricted to bandwidth constraint, we first decouple the multi-user scheduling problem into a single user constrained Markov decision process (CMDP) through relaxation of the hard bandwidth constraint. Next we exploit the threshold structure of the optimal policy for the decoupled single user CMDP and obtain the optimum solution through linear programming (LP). Finally, an asymptotic optimal truncated policy that can satisfy the hard bandwidth constraint is built upon the optimal solution to each of the decoupled single-user sub-problem. The performance is verified through simulations. Our investigation shows that to obtain a small AoI performance, the scheduler exploits good channels to schedule users supported by limited power. Users equipped with enough transmission power are updated in a timely manner such that the bandwidth constraint can be satisfied.
	\end{abstract}
	
	
	\section{Introduction}
	Data freshness is gaining increasing importance in real-time services like real time positioning, monitoring and industrial control. To support these applications, users that track the corresponding physical phenomena are scheduled to send updates to the central controller via time-varying wireless channels. However, the wireless bandwidth and interference constraint, the limited power resource of each user and the time varying nature of wireless channels create obstacles in scheduling strategy design. Moreover, traditional quality of service (QoS) guarantees such as latency and throughput have their limitations and may not guarantee a good data freshness performance. Thus, it is of great importance to revisit sampling and scheduling strategies in wireless networks in order to obtain more fresh information.
	
	A recently proposed metric, the \emph{Age of Information} (AoI) \cite{yates12infocom}, namely the time elapsed since the generation time-stamp of the freshest information stored at the receiver, has been widely adopted to measure data freshness in communication networks. Intuitively, to guarantee a low AoI performance in resource constrained networks, packets with short delay have to be transmitted in a timely manner. Optimizing and analyzing AoI performance in point to point communication systems with power consumption constraint have been studied \cite{yates_15_isit,sun16infocom,sun17TIT,arafa_18_isit,yang_17_isit,ceran_18_wcnc,baknina_18_isit}. In these works, the optimal sampling and transmission strategy in the presence of queueing delay\cite{sun17TIT} and transmission failure \cite{ceran_18_wcnc} are shown to possess a threshold structure, i.e., sampling and update transmission occur when information at the receiver is no longer fresh while the update packets, if successfully received, can significantly reduce data staleness. 
	
	AoI performance and optimization in multi-user network have been studied in \cite{igor16allerton,igor_18_ton,talak2017allerton,talak_18_isit,talak_18_wiopt,hsu17isit,jiang_isit_2018,lu_age_2018,igor18infocom}. When all the users in the network are identical and update packets can be generated at will, a greedy policy that samples and schedules to transmit the user with the largest AoI is shown to be optimal \cite{igor16allerton}. When there is no packetloss in the network, this greedy policy is equivalent to the round robin strategy, which is shown to be order optimal when update packets can not be generated at will and arrive randomly \cite{jiang_isit_2018}. In \cite{igor_18_ton}, it is revealed that users with relatively bad channel states are updated less frequently. Scheduling to minimize AoI performance in networks with time-varying channels are studied in \cite{talak_18_isit,talak_18_wiopt}, where centralized and decentralized policies to minimize the average peak age of information (PAoI) are proposed. However, the channel model considered in these works have two states and no power adaptation strategy is used to combat wireless fading effect.
	
	To combat the aforementioned fading effect, transmission power and bandwidth limitations, which appear at different layers of communication networks, cross-layer control strategies have been studied in \cite{borkar_18_tcns,chen_2018_ton,wang_17_GC,wang_19_tcomm,yang_10_twc,uysal_02_ton,berry_02_tit,singh_2019_tac} to minimize delay or maximize throughput. In \cite{uysal_02_ton}, a Lazy scheduling policy that assigns scheduling decision based on the queue backlog is proposed. Considering time-varying fading nature of wireless channels, rate and power adaptation strategy is proposed in \cite{berry_02_tit}. To minimize queueing delay in a point to point time-varying channel with average power constraint on the transmitter, a probabilistic scheduling strategy is proposed in \cite{wang_17_GC,wang_19_tcomm}.  Although cross-layer strategies have been studied in delay minimization, throughput and utility maximization, the design to optimize Age of Information has not been very well studied.
	
	To fill this gap, in our paper, we consider a single controller schedules multiple users to transmit updates in a wireless network. Similar to the cross-layer framework \cite{collins1999transmission}, the wireless link of each user is modeled to be multi-state time-varying and different level of transmission power is used in different channel state to guarantee success transmission. The overall objective is to minimize the expected average AoI performance when network is restricted to bandwidth constraint and scheduling decisions have to satisfy the power constraint of each user. Inspired by \cite{singh_2019_tac}, we first relax the hard bandwidth constraint and decouple the multi-user scheduling problem into a single user constrained Markov decision process (CMDP). Then we propose a truncated scheduling policy that can achieve an asymptotic average AoI performance over the entire network.
	
	The main contributions of the paper are summarized as follows:
	\begin{itemize}
		\item A cross-layer network opportunistic scheduling framework to study the AoI minimization with power constrained users is proposed. The channel states of all users are assumed to be known at the beginning of each slot through channel estimation before scheduling decisions are made and remain constant during the slot. Different level of transmit power is adopted in different channel state to ensure successful packet transmission. This model captures key features of practical cross-layer network optimization problem and facilitates analysis.
		
		\item By relaxing the hard bandwidth constraint, we decouple the multi-user bandwidth and power constrained scheduling problem into a single-user CMDP. The threshold structure of the optimal policy to the CMDP is exploited and then the CMDP is converted into a Linear Programming (LP).
		
		\item A dual-method is proposed to search for the Lagrange multiplier such that the relaxed bandwidth constraint can be satisfied. Inspired by \cite{singh_2019_tac}, we propose an asymptotic optimal truncated scheduling policy to minimize AoI performance under hard bandwidth constraint. The performance of the algorithm is analyzed and verified through simulations. 
	\end{itemize}
	
	The remainder of this paper is organized as follows. The network model and the data freshness metric, AoI, are introduced in Section II. In Section III, we decouple the multi-user scheduling problem into single-user level CMDP and search for the optimal policy through LP. In Section IV, a truncated multi-user scheduling policy is proposed. Section V evaluates and analyzes the performance of the proposed algorithm and Section VI draws the conclusion. 
	
	\emph{Notations: }Vectors and matrices are written in boldface lower and upper letters. The probability of event $\mathcal{A}$ given condition $\mathcal{B}$ is denoted as Pr$(\mathcal{A}|\mathcal{B})$. The expectation operation with regard to random variable $X$ is denoted as $\mathbb{E}_X[\cdot]$. The cardinality of set $\Omega$ is denoted as $|\Omega|$.
	
	\section{System Model and Problem Formulation}
	\subsection{Network Model}
	We consider a network with a central controller collecting time-sensitive data from $N$ users via wireless links. Let the time be slotted, i.e., $t=\{1, \cdots, T\}$. The central controller schedules users to transmit update at the beginning of each slot over time-varying fading channels. Let the indicator function $u_n(t)$ to be a scheduling decision. If $u_n(t)=1$, then user $n$ is scheduled to transmit update packet during slot $t$ and the packet will be received successfully by the end of the slot. Due to the limited bandwidth resource, no more than $M$ users can be scheduled simultaneously, which casts the following restrictions on $u_n(t)$:
	\begin{equation}
	\sum_{n=1}^Nu_n(t)\leq M, \text{for all }t.
	\end{equation}
	
	We assume that the communication channel between the central controller and each user experiences an independent $Q$-state block fading, where $Q$ is a positive integer. The channel state remains constant during a slot but follows an i.i.d fading process across the slots. Let $q_n(t)\in\{1, \cdots, Q\}$ be a random variable that captures the channel state of user $n$ during slot $t$, large $q_n(t)$ indicates that link $n$ is more noisy and goes through stronger fading during slot $t$. Let the probability mass function of $q_n(t)$ be:
	\begin{equation}
	\text{Pr}(q_n(t)=q)=\eta_{n, q},
	\end{equation}
	where $\eta_{n, q}\in[0, 1]$. For each user $n$, the sum of $\eta_{n, q}$ must satisfy:
	\begin{equation}
	\sum_{q=1}^Q\eta_{n, q}=1.
	\label{etaconstraint}
	\end{equation} 
	
	When user $n$ is scheduled to transmit updates in a slot and the corresponding channel state is $q$, in order to guarantee successful transmission, it will consume $\omega(q)$ units of energy. To combat the effect of channel fading, more power will consumed when the channel is more noisy, thus $\omega(1)<\cdots<\omega(Q)$ is an increasing sequence. The transmitted packet will be successfully received by the central controller at the end of the slot. For a typical scheduling decision $\mathbf{u}_n(\pi)=[u_n(1), \cdots, u_n(T)]$ related to user $n$, the average power consumed in $T$ consecutive slots can be computed as follows:
	\begin{equation}
	E_n(\mathbf{u}_n(\pi))=\frac{1}{T}\sum_{t=1}^Tu_n(t)\omega(q_n(t)).
	\end{equation}
	
	\subsection{Age of Information}
	We measure data freshness of the central controller by using the metric \emph{Age of Information}(AoI)\cite{yates12infocom}. By definition, the AoI is the time elapsed since the generation time-stamp of the freshest information at the receiver.
	
	Let $x_n(t)$ be age of information of user $n$ at the beginning of slot $t$. In this work, it is equivalent to the number of slots elapsed since the last delivery to user $n$. If $u_n(t)=1$, fresh information about user $n$ will be received by the central controller at the end of slot $t$, thus $x_n(t+1)=1$; otherwise, since there is no update packet received from user $n$ during slot $t$, information about user $n$ will be one slot older, hence $x_n(t)$ increases linearly and $x_n(t+1)=x_n(t)+1$. The evolution of AoI $x_n(t)$ is organized as follows:
	\begin{equation}
	x_n(t+1)=\begin{cases}
	1, &u_n(t)=1;\\
	x_n(t)+1, &u_n(t)=0. 
	\end{cases}
	\end{equation}
	
	\subsection{Problem Formulation}
	For a given network setup with $N$ users and channel states distributions $\{\eta_{n,q}\}$, we measure the data freshness of the entire network by following policy $\pi$ in terms of the expected average AoI of all users at the beginning of each time slot for a consecutive of $T\rightarrow\infty$ slots, which is computed as follows: 
	\begin{align}
	J(\pi)=\lim_{T\rightarrow\infty}\{\frac{1}{NT}\mathbb{E}_\pi\left[\sum_{t=1}^T\sum_{n=1}^Nx_n(t)|\mathbf{x}(0)\right]\},
	\end{align}where the vector $\mathbf{x}(t)=[x_1(t), x_2(t), \cdots, x_N(t)]\in\mathbb{N}^N$ denotes the AoI of all users at the beginning of slot $t$. In this work, we assume that all the users have been synchronized initially, i.e., $\mathbf{x}(0)=\mathbf{1}$ and omit it henceforth.
	
	Denote $\Pi_{\text{NA}}$ to be the class of non-anticipated policies, i.e., scheduling decisions are made based on current and past AoI, channel states as well as their probability distributions, while no future information or prediction about channel states are exploited. The central controller is fully aware of the average power constraints of each user and aim at designing policy $\pi\in\Pi_{\text{NA}}$ in order to minimize the average expected AoI of the entire network. The bandwidth and power constrained AoI (B\&P Constrained AoI) minimization problem is organized as follows:
	\begin{pb}[B\&P-Constrained AoI]\label{schedulingprimal}
		\begin{subequations}
			\begin{align}
			\pi^*=\arg&\min_{\pi\in\Pi_{\text{NA}}}\lim\limits_{T\rightarrow\infty}\{\frac{1}{NT}\mathbb{E}_{\pi}\left[\sum_{t=1}^T\sum_{n=1}^Nx_n(t)\right]\},\label{objprimal}\\
			\text{s.t. }&\sum_{n=1}^N u_n(t)\leq M, \forall t,\label{hardinterference}\\
			&\lim_{T\rightarrow\infty}\frac{1}{T}\mathbb{E}_{\pi}\left[\sum_{t=1}^Tu_n(t)\omega(q_n(t))\right]\leq\mathcal{E}_n, \forall n\label{energyconstraint}. 
			\end{align}	
		\end{subequations}
	\end{pb}
	The hard bandwidth constraint (7b) in every slot $t$ suggests that the \emph{B\&P-Constrained AoI} problem is an intractable integer programming problem, which is extremely difficult to handle. We tackle with this challenge through the following approaches:
	\begin{itemize}
		\item Inspired by \cite{singh_2019_tac,chen_2018_ton,yates17isit}, we relax the hard bandwidth constraint (7b) into a time-average constraint and decouple the multi-user scheduling problem into single user CMDP. After relaxation, more than $M$ users can be scheduled simultaneously. 
		\item In Section III-D, the decoupled single user CMDP is solved through LP. And in Section IV, based on the solution to each of the decoupled single user, we propose a truncated scheduling policy that can satisfy the hard bandwidth constraint (7b).
	\end{itemize}
	\section{Scheduling by user-level decomposition}
	In this section, we start by relaxing and decoupling the \emph{B\&P-Constrained AoI} problem, then formulate the decoupled single user scheduling problem into a constrained Markov decision process (CMDP). We exploit the threshold structure of the optimal stationary randomized policy and the optimal solution is solved through linear programming (LP). 
	
	\subsection{Single-User Level Decomposition}
	Let us first relax the hard constraint 
	(7b) into an time-average constraint, the problem of scheduling multiple power constrained users with relaxed bandwidth constraint (RB\&P-Constrained AoI) can be organized as follows:
	\begin{pb}
		[RB\&P-Constrained AoI]\label{relaxed}
		\begin{subequations}
			\begin{align}
			\pi_R^*=\arg&\min_{\pi\in\Pi_{\text{NA}}}\lim\limits_{T\rightarrow\infty}\{\frac{1}{NT}\mathbb{E}_{\pi}\left[\sum_{t=1}^T\sum_{n=1}^Nx_n(t)\right]\},\label{objrelaxed}\\
			\text{s.t. }&\frac{1}{T}\sum_{t=1}^T\sum_{n=1}^N u_n(t)\leq M, \label{relaxedinterference}\\
			&\lim_{T\rightarrow\infty}\frac{1}{T}\mathbb{E}_{\pi}\left[\sum_{t=1}^Tu_n(t)\omega(q_n(t))\right]\leq\mathcal{E}_n, \!\forall n. 
			\end{align}	
		\end{subequations}
	\end{pb}
	Next we establish the Lagrange function and place the relaxed constraint into the objective function \eqref{objprimal} as follows:
	\begin{align}
	\mathcal{L}(\pi, W)\!=\!\lim_{T\!\rightarrow\!\infty}	\{\frac{1}{N\!T}\mathbb{E}_\pi\left[\sum_{n\!=\!1}^N\sum_{t\!=\!1}^T\left(x_n(t)\!+\!W\!u_n(t)\!-\!W\!M\!N\right)\right]\}.
	\label{relaxedopt}
	\end{align}
	
	The Lagrange multiplier $W\geq 0$ associates with the relaxed constraint and can be viewed as a penalty incurred by policies that want to schedule users above the relaxed bandwidth constraint. For fixed $W$, the optimization problem \eqref{relaxedopt} can then be decoupled into $N$ single user cost minimization problem with average power consumption constraint. The objective of user $n$ is to develop a scheduling strategy $\pi_n$ such that under power constraint Eq.~\eqref{energyconstraint}, the average overall cost incurred by AoI and scheduling penalty can be minimized. The decoupled single user power constrained cost minimization problem is organized as follows:
	\begin{pb}
		[Decoupled P-Constrained Cost]\label{singleopt}
		\begin{subequations}
			\begin{align}
			\pi_n^*&=\arg\min_{\pi\in\Pi_{\text{NA}}}\mathcal{L}_n(\pi, W),\nonumber\\
			\text{where }\mathcal{L}_n(\pi, W)&=
			\lim_{T\rightarrow\infty}\frac{1}{T}\mathbb{E}_{\pi}\left[\sum_{t=1}^Tx_n(t)+Wu_n(t)\right],\\
			\text{s.t. }&\lim_{T\rightarrow\infty}\frac{1}{T}\mathbb{E}_\pi\left[\sum_{t=1}^Tu_n(t)\omega(q_n(t))\right]\leq\mathcal{E}_n.
			\end{align} 
		\end{subequations}
	\end{pb}
	Since the primal relaxed problem \eqref{relaxedopt} gets decoupled, we omit the subscript $n$ henceforth. We formulate \emph{Decoupled P-Constrained Cost} minimization problem into an CMDP in Sec. III(B) and analyze the optimum structure in Sec. III(C). In Sec. III(D), we convert the single-user optimization problem with fixed $W$ into a Linear Programming (LP). 
	
	\subsection{Constrained Markov Decision Process Formulation}
	The decoupled single-user scheduling problem can be formulated into a CMDP that consists of a quadruplet $(\mathbb{S}, \mathbb{A}, \text{Pr}(\cdot|\cdot), C(\cdot, \cdot))$, each item is explained as follows:
	\begin{itemize}
		\item \textbf{State Space: }The state of a user at the beginning of slot $t$ is the current number of slots elapsed since the last update and the channel state $(x(t), q(t))$, the state space $\mathbb{S}=\{x\times q\}$ is thus countable but infinite.
		\item \textbf{Action Space: }There are two possible actions $s\in\mathbb{A}=\{0, 1\}$, where $s(t)=1$ denotes updates from the user is scheduled at the beginning of slot $t$, and $s(t)=0$ represents that the user keeps idle and is not scheduled. Notice that $s(t)$ is different from scheduling decision $u(t)$, which has strict bandwidth constraint.
		\item \textbf{Probability Transfer Function: }If the user is not selected to transmit updates in slot $t$, i.e., $s(t)=0$, then the information will be one slot older and AoI increases linearly, $x(t+1)=x(t)+1$, otherwise if the user is scheduled, then $x(t+1)=1$. The channel state $q(t)$ evolves independently of $x(t)$, hence the probability transfer function from state $(x, q)$ is organized as follows:
		\begin{subequations}
			\begin{align}
			&\text{Pr}((x,q)\rightarrow(x+1, q'))=\eta_{q'}, &s=0;\\
			&\text{Pr}((x,q)\rightarrow(1, q'))=\eta_{q'}, &s=1.
			\end{align}
		\end{subequations}
		\item \textbf{One-Step Cost: }For given state $(x, q)$, the one-step cost by taking action $s$ contains AoI growth and scheduling penalty, which can be computed as follows:
		\begin{subequations}
			\begin{equation}
			C_X(x, q, s)=x+Ws,
			\label{onestepcost}
			\end{equation}
			while the one-step power consumption is:
			\begin{equation}
			C_Q(x, q, s)=\omega(q)s.
			\end{equation}
		\end{subequations}
	\end{itemize}
	
	The objective of the decoupled CMDP problem is to design a scheduling policy $\pi$ such that under the average power constraint, \[\frac{1}{T}\mathbb{E}_\pi\left[\sum_{t=1}^TC_Q(x(t), q(t), s(t))\right]\leq \mathcal{E},\]  
	the overall cost containing both AoI and scheduling penalty over infinite horizon can be minimized, which is computed as follows: \[\frac{1}{T}\mathbb{E}_\pi\left[\sum_{t=1}^TC_X(x(t), q(t), s(t))\right].\] 
	\subsection{Characterization of the Optimal Policy}
	In this part, we focus on exploiting the threshold structure of the optimal policy. First we provide the formal definition of a stationary randomized policy and stationary deterministic policy:
	
	\begin{definition}
		Let $\Pi_{\text{SR}}$ and $\Pi_{\text{SD}}$ denote the class of stationary randomized and stationary deterministic policy, respectively. Given observation $(x(t)=x, q(t)=q)$, a stationary randomized policy $\pi_{\text{SR}}\in\Pi_{\text{SR}}$ choose action $s(t)=1$ with probability measure $\xi_{x, q}\in[0, 1]$ for all $t$. A stationary deterministic policy $\pi_{\text{SD}}\in\Pi_{\text{SD}}$ selects action $s(t)=a(x, q)$, where $a(\cdot):(x, q)\rightarrow\{0,1\}$ is a deterministic mapping from state space to action space. 
	\end{definition}
	
	According to \cite[Theorem 4.4]{altman1999constrained}, the optimal policy to the above CMDP has the following property:
	\begin{corollary}
		An optimal stationary randomized policy $\pi^*\in\Pi_{\text{SR}}$ exists for the decoupled single user power constrained scheduling problem \eqref{singleopt}, and it is a mixture of no more than two stationary deterministic policies $\pi_{\text{SD1}}, \pi_{\text{SD2}}\in\Pi_{\text{SD}}$. Let $\lambda$ be the weight of following stationary deterministic policy $\pi_{\text{SD1}}$ and $1-\lambda$ be the weight of following $\pi_{\text{SD2}}$. Then the optimum policy is:
		\begin{equation}
		\pi^*=\lambda\pi_{\text{SD1}}+(1-\lambda)\pi_{\text{SD2}}.
		\label{mixedstrategy}
		\end{equation}
	\end{corollary}
	
	Each of the deterministic policy can be obtained through the Lagrangian primal-dual method \cite{altman1999constrained}. Let $\lambda\geq 0$ be the Lagrange multiplier related to the average power constraint, then the single user CMDP can be converted into an unconstrained MDP, the objective is to minimize the following overall cost by designing policy $\pi$ with no constraint:
	\begin{pb}
		[Decoupled Unconstrained Cost]\label{unconstrainedMDP}
		\begin{align}
		\pi_{ud}^*=\arg\min_{\pi\in\Pi_{\text{NA}}}\lim_{T\rightarrow\infty}\frac{1}{T}\mathbb{E}_\pi[\sum_{t=1}^T(C_X(x(t), q(t), s(t))\notag\\+\lambda C_Q(x(t), q(t), s(t)))]-\lambda\mathcal{E}\nonumber
		\end{align}
	\end{pb}
	
	For given Lagrange multiplier $\lambda$, a stationary deterministic policy to minimize the above unconstrained cost exists. Moreover, there exits a differential cost-to-go function $V(x, q)$ that satisfies the following Bellman equation:
	
	\begin{align}
	V(x, q)+\gamma=\min\{C_X(x, q, 0)+\sum_{q'=1}^Q\eta_{q'}V(x+1, q'),\notag\\ C_X(x, q, 1)+\sum_{q'=1}^Q\eta_{q'}V(1, q')+\lambda C_Q(x, q, 1)\},
	\label{Bellman}
	\end{align}
	where $\gamma$ is the average cost by following the optimal policy. Next, we will prove the threshold structure of the stationary deterministic policy for given $\lambda$, which will present insight for the structure of the optimal stationary randomized policy to solve the CMDP problem \ref{singleopt}. 
	\begin{lemma}
		The optimal stationary deterministic policy for solving the \emph{Decoupled Unconstrained Cost} minimization problem with fixed $\lambda$ possesses a dual threshold structure, which is explained as follows:
		\begin{itemize}
			\item[1] For any channel state $q$, there exists a threshold $\tau_q$, such that when $x\geq\tau_q$, the optimal action $s^*(x, q)=1$ and when $x<\tau_q$, $s^*(x, q)=0$. 
			\item[2] The set of threshold is non-decreasing, i.e., $\tau_1\leq \tau_2\leq\cdots\leq\tau_Q$. 
		\end{itemize}
	\end{lemma}
	
	\begin{IEEEproof}
		The detailed proof is provided in Appendix A. Here we provide an intuitive analysis. Since communication between the user and the controller is power constrained, we only schedule when the information is no longer fresh or the channel state is good, i.e., $x$ is large or $q$ is small. This behavior characterizes a threshold structure. 
	\end{IEEEproof}
	
	\subsection{Probabilistic Scheduling Policy for Single user Case}
	
	Denote $\xi_{x, q}$ to be the probability that the user is scheduled to send updates with age $x$ and channel state $q$. We aim at finding a set of optimal transmission probability $\{\xi_{x, q}^*\}$ such that total cost of AoI performance and scheduling penalty for a single decoupled user can be minimized. From Sec. III(C), a stationary randomized policy that solves \emph{Decoupled P-Constrained Cost} problem is a randomization between two stationary deterministic policies\cite{altman1999constrained}, each of them can be obtained by solving the \emph{Decoupled Unconstrained Cost} minimization problem, which is an unconstrained MDP. Considering the threshold structure of them and Eq.~\eqref{mixedstrategy}, it can be concluded there exists set of non-decreasing thresholds $\tau_q$, for each state $(x, q)$, if $x\geq\tau_q$, the stationary randomized policy is to schedule the user, i.e., $\xi_{x, q}^*=1$. As an outcome, for each of the decoupled single user problem, when $x\geq\tau_Q$, the user will always be scheduled and the AoI $x$ cannot be larger than the largest threshold $\tau_Q$. To find the optimal policy, we choose a large $X_{\text{max}}$ that can guarantee $X_\text{max}\geq \tau_Q$ in the following analysis. 
	
	Denote $\boldsymbol{\mu}=[\mu_1, \mu_2, \cdots, \mu_{X_\text{max}}]^T$ be the steady distribution of the user's AoI, where $\mu_x$ denotes the probability that $x(t)=x$. The probability transfer graph between the states is plotted in Fig. \ref{probabilitytransfer}. Let $\alpha_x$ and $\beta_x$ denote the one step state transition probability from $x(t)=x$ to $x(t+1)=x+1$ and from $x(t)=x$ to $x(t+1)=1$, respectively, i.e.,
	\begin{subequations}
		\begin{align}
		\alpha_x&=\text{Pr}(x(t+1)=x+1|x(t)=x),\label{forward}\\ \beta_x&=\text{Pr}(x(t+1)=1|x(t)=x).\label{backward}
		\end{align}
	\end{subequations}
	From the discussed threshold structure of deterministic policy, with properly selected $X_{\text{max}}$, under the optimal scheduling policy, the steady state distribution $\mu_{X_{\text{max}}}$ will be 0. And we have the following lemma:
	\begin{figure*}[htb]
		\centering
		\includegraphics[width=.7\textwidth]{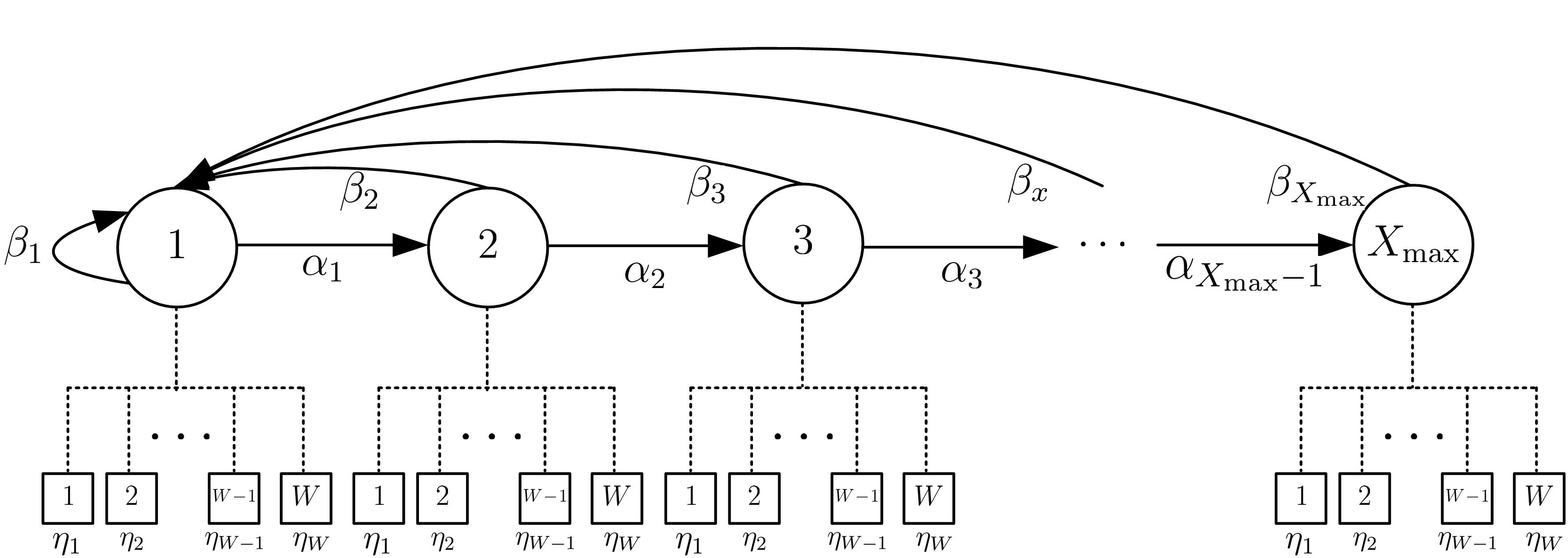}
		\caption{Illustrative of the probability transfer graph for a stationary randomized policy. The circles denotes the AoI and the square denotes the channel states. The forward state transmission probability from AoI $x$ to $x+1$ is $\alpha_x$ and the backward state transmission probability from AoI $x$ to $1$ is $\beta_x$. }
		\label{probabilitytransfer}
	\end{figure*}
	
	\begin{lemma}
		The forward state transfer probability $\alpha_x$ and $\beta_x$ defined in \eqref{forward} and \eqref{backward} can be computed as follows:
		\begin{subequations}
			\begin{align}
			\alpha_x&=\sum_{q=1}^Q\eta_q(1-\xi_{x, q})\\
			\beta_x&=\sum_{q=1}^Q\eta_{q}\xi_q. 
			\end{align}
		\end{subequations}
	\end{lemma}
	\begin{IEEEproof}
		If the state evolves from $x$ to $x+1$, then it can be concluded the user is not scheduled in state $x$. At state $(x, q)$, the probability that the user is not scheduled equals $1-\xi_{x, q}$. According to the law of total probability, 
		\begin{align}
		\text{Pr}(x\rightarrow x+1)&=\sum_{q=1}^Q\sum_{q'=1}^Q\text{Pr}((x, q)\rightarrow (x+1, q'))\eta_q\eta_{q'}\nonumber\\
		&=\sum_{q=1}^Q\sum_{q'=1}^Q\eta_q(1-\xi_{x, q})\eta_{q'}\nonumber\\
		&=\sum_{q=1}^Q\eta_q(1-\xi_{x, q}), \label{alphaproof}
		\end{align}
		where Eq.!\eqref{alphaproof} is obtained because of Eq.~\eqref{etaconstraint}, the sum on $\eta_q$ equals 1. The computation of backward probability $\beta_x$ can be obtained similarly and is hence omitted here. 
	\end{IEEEproof}
	
	Let $\mathbf{Q}$ be the probability transfer matrix between the states, which is,
	\begin{equation}
	\mathbf{Q}=\left[\begin{matrix}
	\boldsymbol{\beta}^T\\
	\mathbf{A},\boldsymbol{0}_{X_\text{max}-1}
	\end{matrix}\right],
	\end{equation} 
	where vector $\boldsymbol{\beta}=[\beta_1, \cdots, \beta_{X_\text{max}}]^T$ is the backward state transition probability vector and $\mathbf{A}=\text{diag}(\alpha_1, \cdots, \alpha_{X_\text{max}-1})$. Vector $\mathbf{0}_{X_\text{max}\!-\!1}$ is a $(X_{\text{max}}\!-\!1)$-dimension vector with all the elements being 0. According to property of the steady state distribution, we have $\mathbf{Q}\boldsymbol{\mu}=\boldsymbol{\mu}$. In addition, considering that $\mu_x=0, \forall x\geq X_{\text{max}}$, then we have $\sum_{x=1}^{X_\text{max}}\mu_x=1$. Thus, the steady distribution $\boldsymbol{\mu}$ relates to strategy $\{\xi_{x, q}\}$ is the solution to the following linear equations:
	\begin{equation}
	\left[\begin{matrix}
	\mathbf{Q}-\mathbf{I}_{X_\text{max}}\\
	\boldsymbol{1}^T_{X_\text{max}}
	\end{matrix}\right]\boldsymbol{\mu}=\left[\begin{matrix}
	\boldsymbol{0}\\1
	\end{matrix}\right],
	\label{steadystatedistribution}
	\end{equation}
	where $\mathbf{1}_{X_{\text{max}}}$ is a $X_{\text{max}}$-dimension column vector with all the elements being 1. 
	

	Next, we will convert the search for the optimal stationary randomized scheduling strategy into an LP. We introduce a new set of variables $y_{x, q}=\mu_x\eta_q\xi_{x, q}$, each denotes the probability of the user is in state $(x, q)$ and is scheduled to transmit an update. With this set of variables, we present the following theorem:
	\begin{theorem}
		The \emph{Decoupled P-Constrained Cost} minimization problem is equivalent to the following LP problem:
		\begin{subequations}
			\begin{align}
			\{y_{x, q}^*\}&=\arg\min_{\{y_{x, q}\}, \{\mu_x\}}\left(\sum_{x=1}^{X_\text{max}}\sum_{q=1}^QWy_{x, q}+\sum_{x=1}^{X_\text{max}}x\mu_x\right),\label{stationaryLPobj}\\
			\text{s.t.  }	&\mu_1=\sum_{x=1}^{X_\text{max}}\sum_{q=1}^Qy_{x, q},\label{stationaryLP1}\\
			&\mu_x= \mu_{x-1}-\sum_{q=1}^Qy_{x-1, q}, \label{stationaryLP2}\\
			&\sum_{x=1}^{X_\text{max}}\mu_x=1, \label{stationaryLP3}\\
			&y_{x, q}\leq \mu_{x}\eta_q, \label{stationaryLP4}\\
			&\sum_{x=1}^{X_\text{max}}\sum_{q=1}^{Q}y_{x, q}\omega(q)\leq\mathcal{E}\label{stationaryLP5}\\
			&0\leq\mu_x\leq 1, 0\leq y_{x, q}\leq 1, \forall x, q.\label{stationaryLP6} 
			\end{align}
			\label{LP}
		\end{subequations}
	\end{theorem}
	
	\begin{IEEEproof}
		Let us compute the equivalent time average cost to Eq.~\eqref{singleopt} by using variables $\{\xi_{x, q}\}$. Given steady state distribution $\boldsymbol{\mu}$, the probability that the user is in state $(x, q)$ is $\mu_x\eta_q$. With probability $\xi_{x, q}$, the user is selected to be scheduled and incurred a cost of $C_X(x, q, 1)=x+W$, and the user is selected to keep idle with probability $1-\xi_{x, q}$ and incurred a cost of $C_X(x, q, 0)=x$. Then the time average cost by following policy $\{\xi_{x, q}\}$ can be computed by:
		\begin{align}
		&\sum_{x=1}^{X_\text{max}}\sum_{q=1}^Q\mu_x\eta_q(\xi_{x, q}(x+W)+(1-\xi_{x, q})x)\nonumber\\
		=&\sum_{x=1}^{X_\text{max}}x\mu_x+\sum_{x=1}^{X_\text{max}}\sum_{q=1}^QWy_{x, q}. 
		\end{align}
		
		For each state $(x, q)$, the power consumed for being active is $\omega(q)$. Then, the time-average power consumed by employing policy $\{\xi_{x, q}\}$ is:
		\begin{equation}
		\sum_{x=1}^{X_\text{max}}\sum_{q=1}^Q\mu_x\eta_q\xi_{x, q}\omega(q)=\sum_{x=1}^{X_\text{max}}\sum_{q=1}^Qy_{x, q}\omega(q),
		\end{equation}
		with this equation the power constraint \eqref{energyconstraint} can be converted in the linear constraint \eqref{stationaryLP5}. The constraint Eq.~\eqref{stationaryLP1}-\eqref{stationaryLP3} can be obtained by substituting $\xi_{x, q}$ with $y_{x, q}$ and $\mu_x$, the relationship is obtained from \eqref{steadystatedistribution}. Notice that $\xi_{x, q}\leq 1$, the inequality constraint \eqref{stationaryLP4} can be obtained. 
	\end{IEEEproof}
	Till now, for fixed Lagrange multiplier $W$, we construct an LP problem to obtain the AoI distribution $\mu_x$ and $y_{x,q}$ by applying the optimum policy to minimize AoI and scheduling cost for a single user. The optimum stationary policy can be computed through $\mu_x$ and $y_{x,q}$. According to the threshold structure of each deterministic policy, we will have the following properties on $\xi^*_{x, q}$:
	\begin{corollary}
		The set of optimal scheduling probabilities $\{\xi_{x, q}^*\}$ possesses the following characteristics:
		\begin{itemize}
			\begin{subequations}
				\item[1] For any channel state $q$:
				\begin{equation}
				\xi_{x_1, q}^*\leq \xi_{x_2, q}^*, \forall 1\leq x_1<x_2.
				\end{equation}
				\item[2] For a specific AoI $x$:
				\begin{equation}
				\xi_{x, q_1}^*\geq \xi_{x, q_2}^*, \forall 1\leq q_1<q_2. 
				\end{equation}
			\end{subequations}
		\end{itemize}
	\end{corollary}
	
	With this corollary, we can then present the threshold structure of the stationary randomized policy:
	\begin{theorem}
		The optimal stationary randomized policy for solving the single-user scheduling problem \eqref{singleopt} under power consumption constraint also possesses a threshold structure, which is explained as follows:
		\begin{itemize}
			\item[1] For any channel state $q$, there exists a threshold $\tau_q$, such that when $x>\tau_q$, it is always optimal to schedule, i.e., $\xi_{x, q}^*=1$ and when $x<\tau_q$, $\xi_{x, q}^*=0$, while the scheduling decision at the $\tau_q$ may be a randomized strategy, i.e., $0<\xi_{\tau_q, q}\leq1$. 
			\item[2] The set of threshold is non-decreasing, i.e., $\tau_1\leq \tau_2\leq\cdots\leq\tau_Q$. 
		\end{itemize}
	\end{theorem}
	\begin{IEEEproof}
		Suppose $\{\xi_{x, q}^*\}$ is the set of optimum decision and $(\boldsymbol{\mu}^*, \mathbf{y}^*)$ is the optimizer to the LP. If $\sum_{x=1}^{X_\textbf{max}}\sum_{q=1}^Qy_{x, q}\omega(q)<\mathcal{E}$, suggesting the average power consumed by the optimum scheduling policy for the decoupled single user didn't meet the power constraint. Then it can be concluded the solution to the single user CMDP is the same to the decoupled single user problem (P\eqref{singleopt}) without the power consumption constraint (10b). Since optimal solution to unconstrained MDP belongs to the class of stationary deterministic policy, it can be concluded that $\xi_{x, q}^*=\{0, 1\}$. Then considering Corollary 2, sequence $\xi_{\cdot, q}^*$ is increasing for fixed $q$ and sequence $\xi_{x, \cdot}^*$ is increasing for fixed $x$. Then Theorem 2 can be verified. 
		
		If $\sum_{x=1}^{X_\textbf{max}}\sum_{q=1}^Qy_{x, q}\omega(q)=\mathcal{E}$, suggesting the optimum policy uses up all the power constraint. Notice that Corollary 2 is similar to \cite[Lemma 2]{wang_19_tcomm} and the proof for Theorem 2 in this case can be carried out in a similar manner to \cite[Theorem 5]{wang_19_tcomm}, then it can be concluded that for all the $\{\xi_{x, q}\}$, there exists at most one state $(x', q')$ such that $\xi_{x', q'}\in(0, 1)$ and for all the other states $(x, q)\neq(x', q')$, the scheduling probability $\xi_{x, q}\in\{0, 1\}$. Thus, considering Corollary 2, the threshold structure of the optimum stationary randomized policy and the increasing characteristic of the thresholds $\tau_q$ can be verified. 
	\end{IEEEproof}
	
	%
	
	\section{Multi-user Opportunistic Scheduling}
	
	In this section, we will provide an algorithm to determine the multiplier $W$ such that relaxed bandwidth constraint can be satisfied. Then, we propose an asymptotic optimal truncated scheduling algorithm for the multi-user case that satisfies the original hard bandwidth constraint Eq.~\eqref{hardinterference}. 
	\subsection{Determination of Lagrange Multiplier}  
	After solving the single user problem for fixed $W$, by combining the optimum scheduling strategy $s_n(t)$ for each of the user, the optimal policy $\pi_R^*(W)$ to minimize the Lagrange function Eq.~\eqref{relaxedopt} for fixed $W$ can be obtained. Next, we describe how to obtain the optimal Lagrange multiplier $W$ so that the \emph{RB\&P-Constrained AoI} problem can be solved.
	
	Let $g(W)$ denote the Langrangre dual function, i.e., 
	\begin{equation}
	g(W)=\min_{\pi\in\Pi_{\text{NA}}}\mathcal{L}(\pi, W).
	\end{equation}
	
	Since the relaxed problem gets decoupled into $N$ single user CMDP, the dual function can be computed by:
	\begin{align}
	g(W)=\frac{1}{N}\sum_{n=1}^N&g_n(W)-WM, \text{where }\notag\\g_n(W)=\min_{\pi_n\in\Pi_{\text{NA}}}&\left(\mathcal{L}_n(\pi_n, W)\right), \text{ s.t. } \text{Eq. ~(7c)}.
	\end{align}
	
	By Theorem 1, the CMDP that minimizes $\mathcal{L}_n(\pi, W)$ is equivalent to an LP, then $g_n(W)$ equals the average cost of the CMDP\cite{altman1999constrained}. Let $\overline{X}_n(W)$ and $\overline{A}_n(W)$ denote the average AoI and the average activation probability for user $n$, respectively. By computing the optimum resource allocation vector $\{y_{x, q}^*\}$ through solving LP \eqref{LP}, the dual function $g_n(W)$ can be computed as follows:
	\begin{subequations}
		\begin{align}
		g_n(W)=\overline{X}_n(W)&+W\overline{A}_n(W)\\
		\text{where } \overline{X}_n(W)&=\sum_{x=1}^{X_{\text{max}}}x\mu_x^*,\\
		\overline{A}_n(W)&=\sum_{x=1}^{X_\text{max}}\sum_{q=1}^Qy_{x, q}^*.
		\end{align}
	\end{subequations}
	
	Finally, we apply the subgradient descent method to search for the Lagrange optimizer. Let $W^{(k)}$ be the Lagrange multiplier used in the $k^{\text{th}}$ iteration. According to \cite[Eq.~6.1.1]{bertsekas2015convex}, the subgradient at $W^{(k)}$ can be computed by:
	\begin{align}
	\text{d}_{W}g( W^{(k)})&=\sum_{n=1}^N\overline{A}_n( W^{(k)})-M.
	\end{align}
	
	We start with $W^{(0)}=0$, if $\sum_{n=1}^N\overline{A}_n(W^{(0)})-M\leq 0$, then scheduling does not have to consider the relaxed bandwidth constraint. Otherwise, we adopt an iterative algorithm update. By choosing a set of stepsizes $\gamma_k$ similar to \cite{ceran_18_wcnc}, the multiplier for the next iteration can be computed by: 
	\begin{align}
	W^{(k+1)}=W^{(k)}+\gamma_k\text{d}_Wg( W^{(k)}).
	\end{align}
	
	The iteration ends until $|W^{(k)}-W^{(k+1)}|<\varepsilon$. 
	
	However, since the \emph{RB\&P-Constrained AoI} is a constrained Markov decision process, the optimum scheduling policy of which should be a randomization between no more than two policies, each is the solution to minimize the Lagrange function Eq.~\eqref{relaxedopt}. The randomization between the two policies will enable us to satisfy the relaxed bandwidth constraint Eq.~\eqref{relaxedinterference} in the \emph{RB\&P-Constrained AoI}. Next, we will talk about how to obtain the optimum randomized strategy from the obtained Lagrange multipliers sequence $\{W^{(k)}\}$.
	
	Let $W_l$ and $W_u$ be two Lagrange multipliers chosen from sequence $W^{(k)}$, 
	\begin{subequations}
		\begin{align}
		W_l=\arg\max_{W^{(k)}}\sum_{n=1}^N\overline{A}_n(W^{(k)}),\text{s.t. }\sum_{n=1}^N\overline{A}_n(W^{(k)})\!\leq\! M,\\
		W_u=\arg\min_{W^{(k)}}\sum_{n=1}^N\overline{A}_n(W^{(k)}),\text{s.t. }\sum_{n=1}^N\overline{A}_n(W^{(k)})\!\geq\! M.
		\end{align}
	\end{subequations}
	
	Then, let $M_l$ and $M_u$ be the total bandwidth used with respect to minimize the function Eq.~\eqref{relaxedopt}. Let $\{\boldsymbol{\mu}^{n, l},\mathbf{y}^{n, l}\}$ be solution to user n's LP problem (\ref{LP}) with multiplier $W_l$ and $\{\boldsymbol{\mu}^{n, u},\mathbf{y}^{n, u}\}$ is the solution with multiplier $W_u$. To satisfy the relaxed bandwidth constraint, the optimum distribution $\{\boldsymbol{\mu}^{n,*}, \mathbf{y}^{n, *}\}$ of the relaxed problem is a linear combination of $\{\boldsymbol{\mu}^{n, l},\mathbf{y}^{n, l}\}$ and $\{\boldsymbol{\mu}^{n, r},\mathbf{y}^{n, r}\}$, which can be computed as follows:
	\begin{equation}
	\{\boldsymbol{\mu}^{n,*}, \mathbf{y}^{n, *}\}=\lambda\{\boldsymbol{\mu}^{n, l},\mathbf{y}^{n, l}\}+(1-\lambda)\{\boldsymbol{\mu}^{n, u},\mathbf{y}^{n, u}\},
	\end{equation}
	where the coefficient $\lambda$ can be computed in a similar manner to \cite{ceran_18_wcnc}:
	\[\lambda=\frac{M_u-M}{M_u-M_r}.\]
	Notice that $\{\boldsymbol{\mu}^{n,*}, \mathbf{y}^{n, *}\}$ still satisfy the constraint of the LP problem for user $n$. Consider the structure of each \emph{Decoupled P-Constrained Cost} problem, the optimum scheduling strategy $\pi_R^*$ for the \emph{RB\&P-Constrained} is then constructed as follows:
	
	In each slot $t$, the central controller observes the current AoI $x_n(t)$ and channel state $q_n(t)$ of user $n$, a scheduling decision $s_n(t)=1$ is then made with probability $\xi_{x_n(t), q_n(t)}^{n, *}$ is can be computed as follows:
	\begin{equation}
	\xi_{x, q}^{n}=\begin{cases}
	1, &\xi_{x-1, q}^{n}=1\text{ or }\mu_x^{n, *}=0\text{ or }x\geq X_{\text{max}};\\
	\frac{y_{x, q}^{n, *}}{\mu_x^{n, *}\eta_q}, &\text{otherwise}.
	\end{cases}
	\label{SRdistribution}
	\end{equation}
	
	Finally, the minimum AoI performance to the \emph{RB\&P-Constrained AoI} problem can be computed through according to the optimizer $\{\boldsymbol{\mu}^{n, *}, \mathbf{y}^{n, *}\}$, which also formulates the lower bound on the AoI performance to the primal \emph{B\&P-Constrained AoI}:
	\begin{equation}
	\text{AoI}_{\text{LB}}=\text{AoI}_\text{R}^*=\sum_{n=1}^N\sum_{x=1}^{X_\text{max}}x\mu_{x}^{n, *}.
	\end{equation}
	
	\subsection{Multi-user opportunistic scheduling with hard constraint}
	In this part we construct a truncated policy $\pi$ based on optimal scheduling policy for each of the decoupled user and solve the primal \emph{B\&P-Constrained AoI} problem. Let $\pi_R^*$ be the optimum scheduling policy obtained in Sec. IV(A), where $s_n(t)$ is the scheduling decision under the relaxed constraint, which measures if user $n$ is eager be scheduled. Denote $\Omega(t)=\{n|s_n(t)=1\}$ as the set of users that are eager to be scheduled. The scheduling decision $u_n(t)$ under hard bandwidth constraint is then carried out as follows:
	\begin{itemize}
		\item If $|\Omega(t)|\leq M$, i.e., the total number of users that are eager to send updates is less than the available bandwidth, then all the users that are eager to be scheduled can send their updates, i.e., $u_n(t)=1, \forall s_n(t)=1$. \item Otherwise if $|\Omega(t)|>M$, the central controller selects a subset of $\mathcal{M}(t)=|M|\in\Omega(t)$ users from $\Omega(t)$ and schedule them to send updates. Those users that are in set $\Omega(t)$ but not selected in $\mathcal{M}(t)$ is not selected because of limited bandwidth constraint. 
	\end{itemize}
	
	\begin{theorem}
		With the proportion of scheduling resources $\frac{M}{N}=\theta$ keeps a constant, the deviation from the optimal scheduling policy for a network with $N$ users under the proposed truncated policy $\tilde{\pi}$ is $\mathcal{O}(\frac{1}{\sqrt{N}})$. Thus, with $N\rightarrow\infty$ and $\frac{M}{N}=\theta$, the proposed truncated policy is shown to be asymptotically optimal for the primal \emph{B\&P-Constrained AoI} problem with hard bandwidth constraint.
	\end{theorem}
	\begin{IEEEproof}
		The detailed proof is provided in Appendix C.
	\end{IEEEproof}
	
	\section{Simulations}
	In this section, we provide simulation results to demonstrate the performance of the proposed scheduling policy. Notice that from \cite{igor_18_ton}, the optimal policy to minimize AoI performance when all the users are identical is a greedy policy that selects the user with the largest AoI. If there is no packet loss in the network, the greedy policy is equivalent to round robin, which requires a minimum power consumption of $\mathcal{E}_n^{\text{RR}}=\frac{M}{N}\sum_{q=1}^Q\eta_{n, q}\omega(q)$ for user $n$. In the following simulations, we measure power consumption constraint through ratio $\rho_n=\mathcal{E}_n/\mathcal{E}_n^{\text{RR}}$. Small $\rho$ indicates that the corresponding user has a smaller amount of average power budget. We consider a $Q\!=\!4$ states time-varying channel, the distribution is assumed to be $\eta=[0.135, 0.239, 0.232, 0.394]$ and $\omega(q)=q$ for all users. All the simulation results are obtained over a consecutive of $T=10^6$ slots.

\begin{figure}[h]
		\centering
		\includegraphics[width=.48\textwidth]{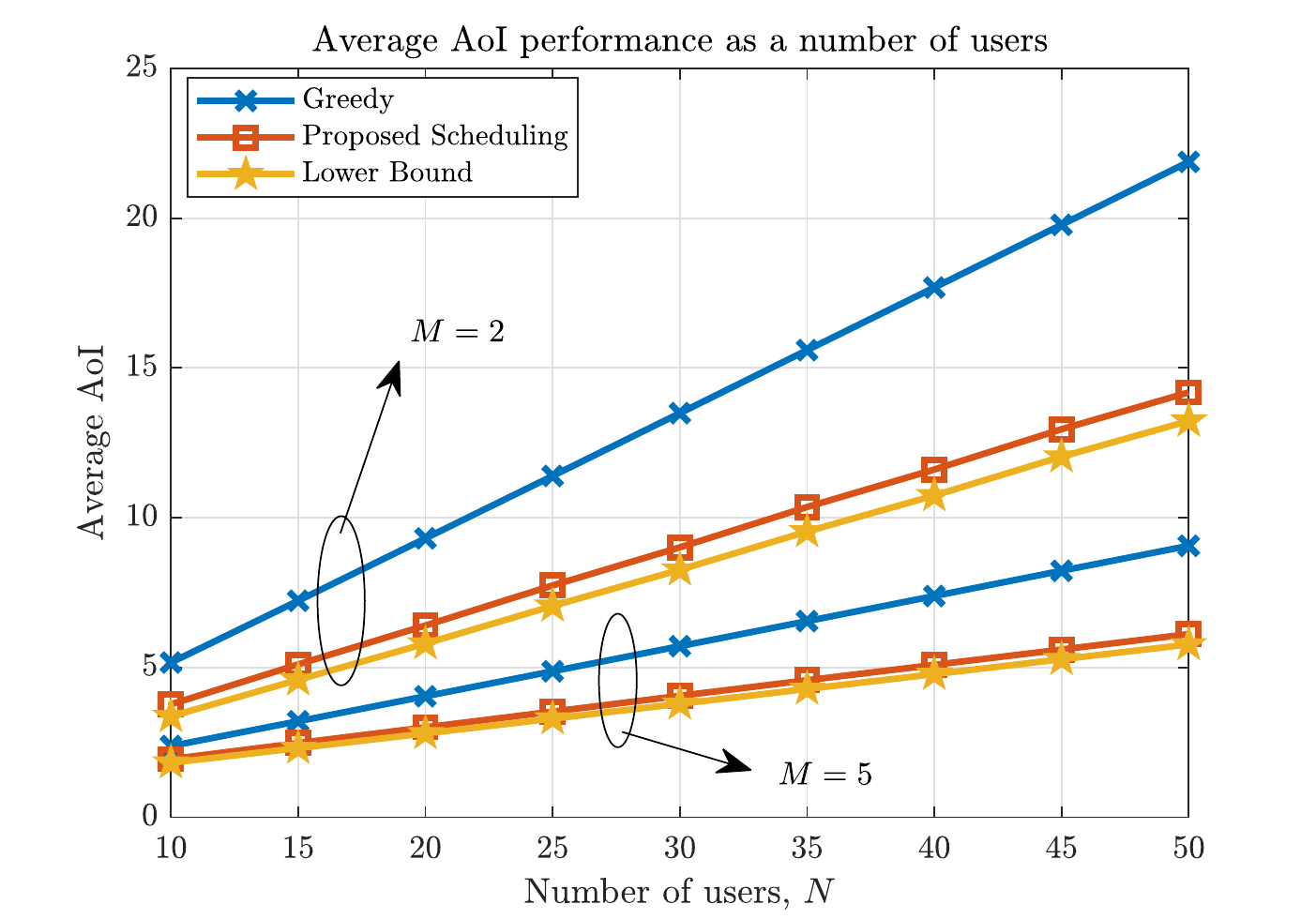}
		\caption{Average AoI performance as a number of users $N$.}
		\label{AoIn}
	\end{figure}

	Fig. \ref{AoIn} studies average AoI performance as a number of users, $N=\{10, 15, \cdots, 50\}$. The power constraint factor is taken from $[0.2, 1.6]$, i.e., $\rho_n=0.2+\frac{1.4}{N}(n-1)$ and the bandwidth $M=\{2, 5\}$. Denote $C_n(t)$ as the total power consumed by user $n$ until slot $t$ and let $\mathcal{R}(t)=\{n|\mathcal{E}_nt-C_n(t)\geq0\}$ be the set of users that has enough power to support transmission in slot $t$. We compare the proposed policy with a naive greedy policy that selects no more than $M$ users with the largest AoI from set $\mathcal{R}(t)$ for scheduling. As can be seen from the figure, the proposed truncated scheduling achieves a close average AoI performance to the lower bound and can achieve more than 30\% AoI decrease compared to the greedy algorithm when $N=50$.
	
\begin{figure}[h]
		\centering
		\includegraphics[width=.48\textwidth]{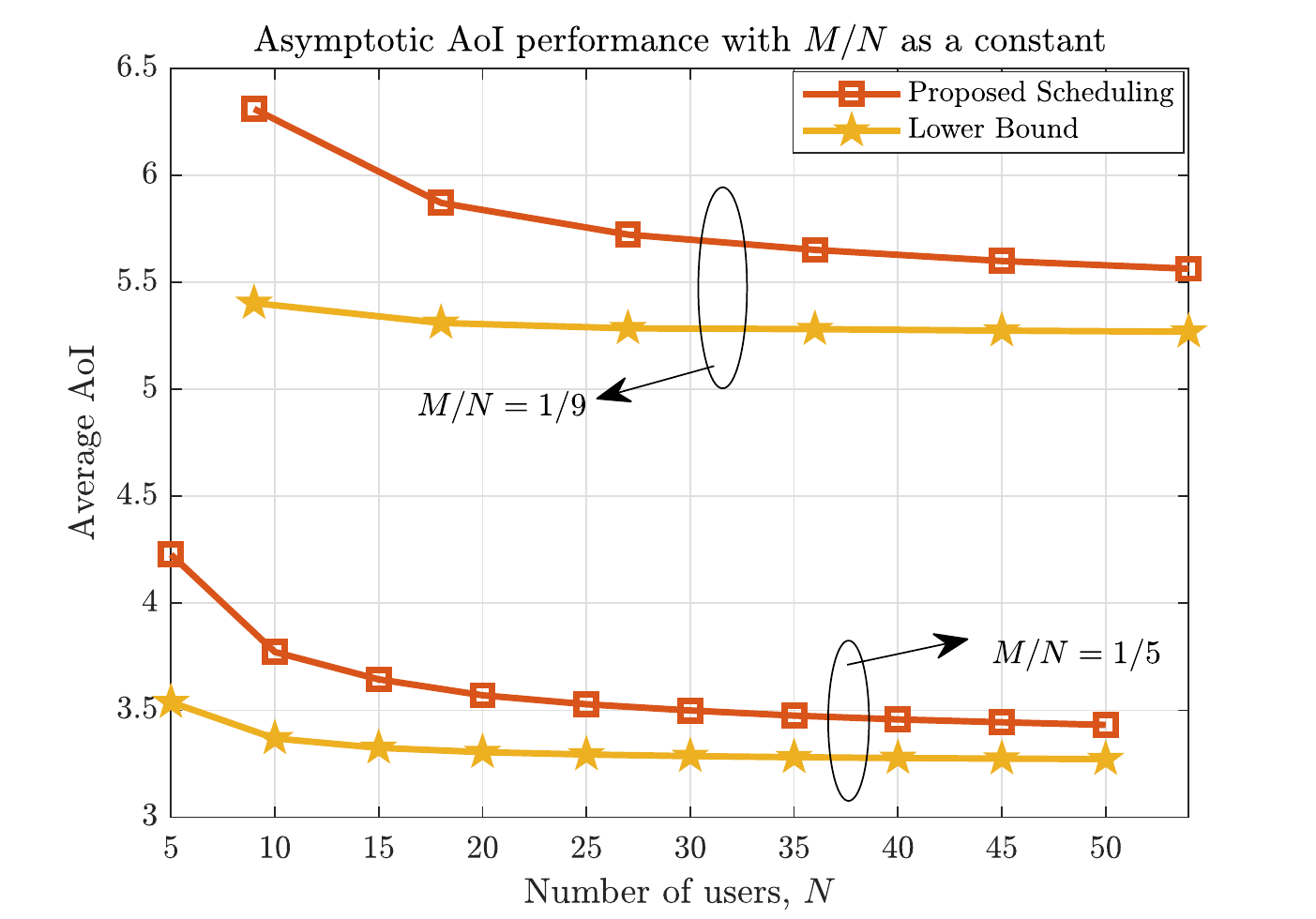}
		\caption{Asymptotic average AoI performance as a number of users $N$, available bandwidth is chosen by $M/N=\{\frac{1}{5}, \frac{1}{9}\}$.}
		\label{AoInasymptotic}
	\end{figure}

	Fig. \ref{AoInasymptotic} studies the asymptotic average AoI performance as a number of users, with $\frac{M}{N}=\{\frac{1}{5}, \frac{1}{9}\}$. The power constraint of each user is selected by $\rho_n=0.2+\frac{1.4}{N}(n-1)$. As can be observed from the figure, the difference between the proposed strategy and the lower bound decreases with $N$. The asymptotic performance is also verified in simulation results.
	
\begin{figure*}[t]
		\centering
		\includegraphics[width=.8\textwidth]{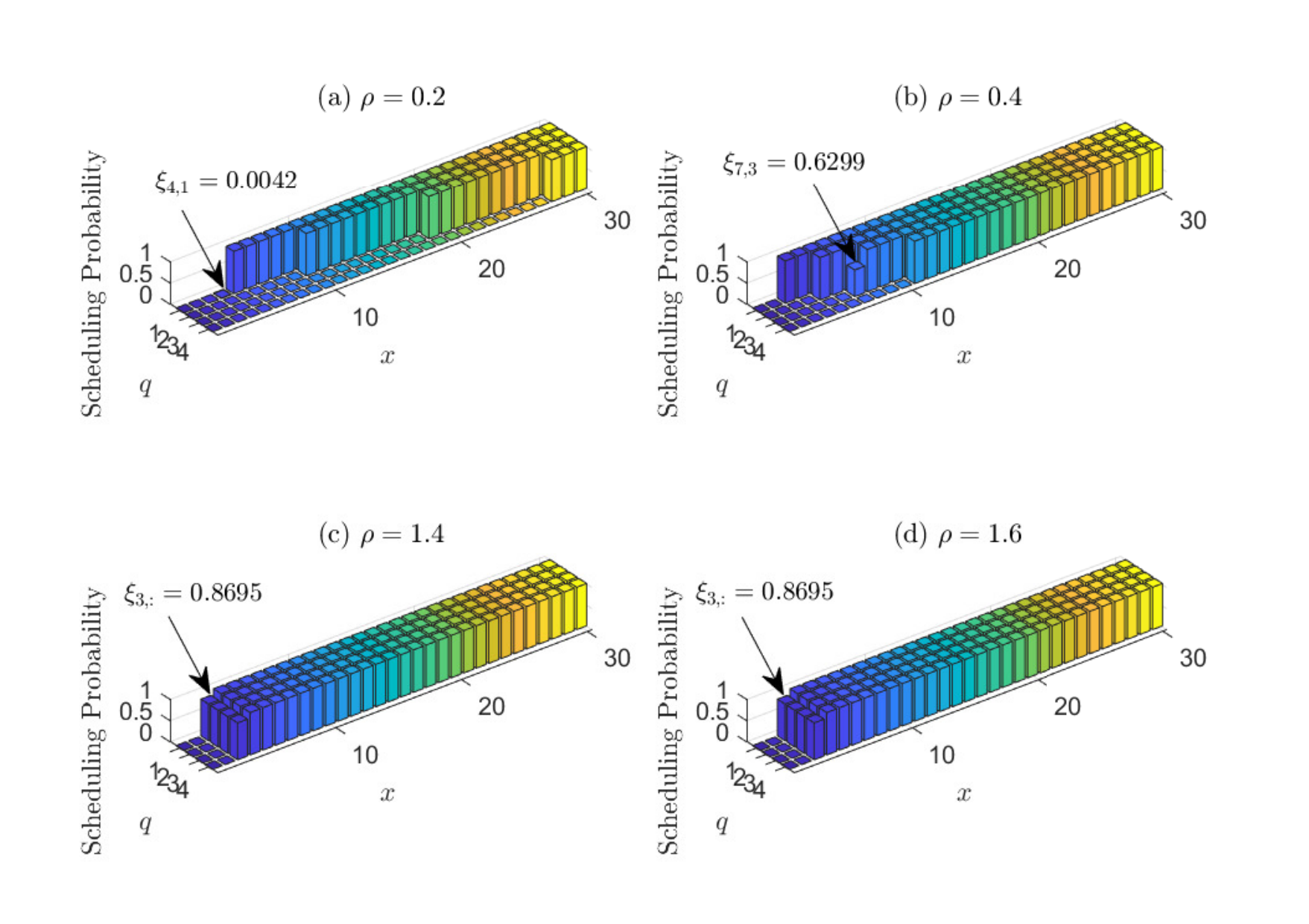}
		\caption{Scheduling decisions for users with different power constraint $\rho_n$. }
		\label{stategyplot}
	\end{figure*}

	
	We visualize the scheduling policy for some representative users in Fig. \ref{stategyplot}. The network consists of $N=8$ users with bandwidth $M=2$, the power constraint factor for each user is $\rho_n=0.2n$. Fig. (a)-(d) demonstrate user $\{1, 2, 7, 8\}$ with power constraint $\rho=\{0.2, 0.4. 1.4, 1.6\}$, respectively. In Fig. \ref{stategyplot}(a)-(b), the transmission power for each user is limited, the scheduling threshold $\tau_q$ is a increasing sequence as channel state $q$. Moreover, the threshold of each channel stated in Fig. \ref{stategyplot}(b) is smaller than corresponding threshold in Fig. \ref{stategyplot}(a), indicating that transmission is more likely to happen as a result of more available transmission power. The optimal strategy for a single user seeks to exploit a good channel state in order to satisfy the power constraint, while trying to keep the AoI small. If unfortunately the channel state is always bad, he will keep waiting until data staleness cannot be bare anymore or the channel state turns good. By comparing Fig.~\ref{stategyplot}(a)-(b), the scheduler tries to make full use of the transmission power through a refinement of activation thresholds. When $\rho=1.4$, power consumption is not a problem, all the channel states shares identical activation threshold that can satisfy the relaxed bandwidth constraint, a user cannot be selected to send updates all the time even if he has enough power. Considering the greedy AoI performance depicted in Fig.~\ref{AoIn}, although greedy algorithm attempts to use up the power and bring smaller AoI performance to users equipped with enough power, it fails to exploit a good channel and opportunistically schedules those power constrained users, hence lead to a much higher AoI performance. Thus, for a network with different power constrained users, the scheduling strategy for different users varies according to their power constraints. The scheduler seeks good channels to carry out scheduling decisions for those power constrained users, while users supported by enough power are updated in a timely manner that can satisfy bandwidth constraint. 
	
	\section{Conclusions}
	
	In this work, we investigate into the problem of age minimization scheduling in power constrained wireless networks, where communication channels are multi-state time varying and different levels of transmission power is adopted to ensure successful transmission. We decouple the multi-user scheduling problem into a single user level constrained Markov decision process. We reveal the threshold structure of the optimal stationary randomized policy for the single user and convert the optimal scheduling problem into a linear programming. An asymptotic optimal truncated scheduling policy for multi-user scenario that satisfies the hard bandwidth constraint is proposed. It is revealed that when power of the user is very limited, the scheduler seeks to exploit a good channel state while keeping the information fresh and minimize the scheduling opportunities. Users equipped with sufficient power are updated in a timely manner that can satisfy the hard bandwidth constraint. 
	
\appendices
\section{Proof of Lemma 1}
\begin{IEEEproof}
	The threshold structure of the optimal policy that minimizes the average cost of \eqref{unconstrainedMDP} is proved by insights from the $\alpha$-discounted cost problems, where $0<\alpha<1$ is a discount factor. Given state $(x, q)$, the expected $\alpha$- discounted cost starting from the state over infinite horizons by following policy $\pi$ can be computed:
	\begin{align}
	J_{\alpha,\pi}(x, q)=\lim_{T\rightarrow\infty}\mathbb{E}_\pi\left[\sum_{t=0}^T\alpha^t[C_X(x(t), q(t), s(t))\nonumber\right.\\
	\left.+\lambda C_Q(x(t), q(t), s(t))]|(x(0)=x, q(0)=q)\right]
	\end{align}
	Let $V_{\alpha}(x, q)=\min_{\pi\in\Pi_{\text{NA}}}J_{\alpha,\pi}(x, q)$ be the minimum expected total discounted cost starting from state $(x, q)$. Then, the minimum total discounted cost will satisfy the following equation:
	\begin{align}
	V_\alpha(x, q)=\min\{C_X(x, q, 0)+\alpha\sum_{q'=1}^Q\eta_{q'}V_\alpha(x+1, q'),\notag\\ C_X(x, q, 1)+\lambda C_Q(x, q, 1)+\alpha \sum_{q'=1}^Q\eta_{q'}V_\alpha(1, q')\},
	\label{discountBellman}
	\end{align}
	
	To verity the threshold structure of the optimal policy to the total discounted cost problem, we will introduce the following characteristic of $V_{\alpha}(x, q)$:
	
	\begin{lemma}
		For given discount factor $\alpha$ and fixed channel state $q$, the value function $V_\alpha(\cdot, q)$ increases monotonically with $x$; for fixed $x$, the value function $V_\alpha(x, \cdot)$ is a non-decreasing function of channel state $q$.
	\end{lemma}
	
	The details of the proof will be given in Appendix B. Finally, let us verify the threshold structure.
	
	(1). \emph{For any channel state $q$, there exists a threshold $\tau_q$, such that the optimal action $s_\alpha^*(x)=1, \forall x\geq\tau_q$ and $s_\alpha^*(x)=0, \forall x<\tau_q$.}

	If the optimal policy $s_\alpha^*(x, q)=1$, i.e, it is better to schedule the sensor at state $(x, q)$, then we can obtain the following inequality because of Bellman equation:
	\begin{align}
	C_X(x, q, 0&)+\alpha\sum_{q'=1}^Q\eta_{q'} V_\alpha(x+1, q')\\
	\geq &C_X(x, q, 1)+\lambda C_Q(x, q, 1)+\alpha\sum_{q'=1}^Q\eta_{q'}V_\alpha(1, q'). \nonumber
	\end{align}By substituting Eq.~\eqref{onestepcost}
	into the Bellman equation, we will have the following inequality of the value function:
	\begin{equation}
	\alpha\sum_{q'=1}^Q\eta_{q'}V_\alpha(x+1, q')\geq (\lambda\omega(q)+W)+\alpha\sum_{q'=1}^Q\eta_{q'}V_\alpha(1, q').
	\label{activatecondition}
	\end{equation}
	According to Lemma 3, the value function $V_\alpha(\cdot, q)$ is monotonic increasing. Hence, for any $x'>x$, we have the following inequality:
	\begin{align}
	\alpha\sum_{q'=1}^Q\eta_{q'}V_\alpha(x'+1, q')>\sum_{q'=1}^Q\eta_{q'}V_\alpha(x+1, q')\nonumber\\
	\geq\alpha(\lambda\omega(q)+W)+\sum_{q'=1}^Q\eta_{q'}V_\alpha(1, q'),
	\end{align}
	which implies that for state $x'>x$, the optimal policy for state $(x', q)$ is to schedule the user. If at state $(x, q)$ the optimal policy is to be passive, then for state $x'<x$, the optimal policy satisfies $s^*(x', q)=0$ can be verified similarly. 
	
	Moreover, for any state $q$, according to the Bellman equation, the expected total discounted cost for keeping idle equals $C_X(x, q, 0)+\alpha\sum_{q'=1}^Q\eta_{q'}V_\alpha(x+1, q')\geq x+\alpha(x+1)$ , which increases linearly with $x$. Hence for any channel state, there must be some state such that inequality \eqref{activatecondition} is satisfied. This suggests that the optimal solution cannot keep passive all the time. Thus, there exists a threshold $\tau_q$ for any state $x>\tau_q$, the optimal policy $s_\alpha^*(x, q)=1$ and for state $x<\tau_q$, $s_\alpha^*(x, q)=0$. The first characteristic is hence verified.
	
	(2) \emph{The threshold $\tau_q$ is a non-decreasing sequence.}
	
	Notice $\omega(q)$ is an increasing sequence, then if $s_\alpha^*(x, q)=1$, for any better channel state $\tilde{q}<q$, we will have the following inequality:
	\begin{align}
	\alpha\sum_{q'=1}^Q\eta_{q'}V_\alpha(x+1, q')\geq (\lambda\omega(q)+W)+\alpha\sum_{q'=1}^Q\eta_{q'}V_\alpha(1, q')\nonumber\\
	>(\lambda\omega(\tilde{q})+W)+\alpha\sum_{q'=1}^Q\eta_{q'}V_\alpha(1, q').
	\end{align}
	
	Thus, it is always optimal to schedule and transmit update when the user is at a better channel state than $q$, then the threshold $\tau_{\tilde{q}}\leq \tau_q$. As a result, it can then be concluded that $\tau_1\leq\cdots\leq\tau_Q$ and the second characteristic can be verified.
	
	Finally, we present the generation of the threshold structure for total discounted cost to establish the structure of the average cost. Take a sequence of discount factors such that $\lim_{k\rightarrow\infty}\alpha_k=1$. Then according to \cite{senottdiscount}, the optimal policy $s_{\alpha_k}^*$ for minimizing the total $\alpha_k$-discounted cost converges to the policy for minimizing the time-average cost, which verifies the threshold structure of the optimal policy $s^*$ as stated in Lemma 1. 
\end{IEEEproof}

\section{Proof of Lemma 3}
\begin{IEEEproof}
	In this section, we aim at verifying the monotonic characteristic of the discounted value function. The value of $V_\alpha(x, q)$ can be computed through value iteration regarding the Eq.~\eqref{discountBellman}. Denote $V_\alpha^{(k)}(x, q)$ to be the value function obtained after the $k^{\text{th}}$ iteration, the monotonic characteristic if verified by induction. 
	
	Suppose $V_\alpha^{(k)}(\cdot, q)$ is non-decreasing. With no loss of generality, suppose $x_1<x_2$. According to the one step cost, we have:
	\begin{equation}C_X(x_1, q, s)<C_X(x_2, q, s), C_Q(x_1, q, s)=C_Q(x_2, q, s).
	\end{equation}
	
	Denote $J_{\alpha,0}^{(k)}(x, q)$ to be the expected total discounted cost if take action $s$. Then we have the following inequality:
	\begin{align}
	&J_{\alpha, 0}^{(k)}(x_1, q)\nonumber\\
	=&C_X(x_1, q, 0)+\sum_{q'}\eta_{q'}V_\alpha^{(k)}(x_1+1, q)\nonumber\\
	\overset{(a)}{<}&C_X(x_2, q, 0)+\sum_{q'}\eta_{q'}V_\alpha^{(k)}(x_2+1, q)\nonumber\\
	=&J_{\alpha, 0}^{(k)}(x_2, q),
	\end{align}
	where inequality (a) is obtained because of the monotonic characteristic of $V_\alpha^{(k)}(\cdot, q)$. Similarly, we will have the conclusion that $J_{\alpha, 1}^{(k)}(x_1, q)<J_{\alpha, 1}^{(k)}(x_2, q)$. Notice that the value function obtained in the $(k+1)^{\text{th}}$ iteration is obtained by:
	\[V_\alpha^{(k+1)}(x, q)=\min_{s}J_{\alpha, s}^{(k)}(x, q),\]
	and for any $s$, $J_{\alpha, s}^{(k)}(x_1, q)<J_{\alpha, s}^{(k)}(x_2, q)$. Thus, the value function $V_{\alpha}^{(k+1)}(x_1, q)<V_{\alpha}^{(k+1)}(x_2, q)$. By letting $k\rightarrow\infty$, the value function $V_\alpha^{(k)}(x, q)\rightarrow V_{\alpha}(x, q)$. Hence, $V_{\alpha}(\cdot, q)$ is monotonic increasing. The characteristic that $V_{\alpha}(x, \cdot)$ is non-decreasing can be verified similarly and is hence omitted. 
\end{IEEEproof}

\section{Proof of Theorem 3}
	\begin{IEEEproof}
		Denote $\pi_R^*$ be the policy that in each slot, schedule all the users with $s_n(t)=1$ and let $\tilde{\pi}$ be the truncated policy described in Sec. V(B). Since $\pi_R^*$ is the optimum performance to the \emph{RB\&P-Constrained AoI} problem, which formulates the lower bound on the primal \emph{B\&P-Constrained AoI} problem. We verify the asymptotic optimality of the proposed scheduling algorithm by computing the expected AoI difference obtained by $\pi_R^*$ and $\tilde{\pi}$. 
		
		First, considering that $\pi_R^*$ satisfy the relaxed constraint, the average number of users that are eager to send updates by following policy $\pi_R^*$ can then be bounded:
		\begin{equation}
		\overline{\Omega}=\mathbb{E}[|\Omega(t)|]\leq M\label{OmegaUB}.
		\end{equation}
		
		According to Theorem 2, the optimum policy to each decoupled single-user optimization problem possesses a threshold structure. Let $\Gamma_n=\tau_{n, Q}-\tau_{n,1}$ be the difference between the activation threshold of user $n$ in channel state $Q$ and channel state $1$. Suppose in slot $t$, $s_n(t)=1$ but user $n$ is not scheduled. If in the next slot, the channel evolves into a worse state with probability $p$, then the user is likely not to be scheduled; or in the next slot although the channel is the same of better than the previous one, there are still more than $M$ users that eager to be scheduled and only $M$ users can be selected randomly. According to the previous statement, the probability that the user is not scheduled in the next slot can be computed by:
		\[p+(1-p)\frac{N-M}{N}=\frac{N-M}{N}+\frac{M}{N}p.\] 
		Notice that the probability that the channel evolves into a worse state is smaller or equal to $1-\eta_{n, 1}$. Thus the probability of not scheduled in the next slot can be upper bounded by $z=\frac{N-M}{N}+\frac{M}{N}(1-\eta_{n, 1})$. As a result, if using policy $\pi_R$, sensor $i$ should be scheduled at time $t$. Then, the probability that sensor $n$ is still not scheduled in the next $t'$ slots by policy is upper bounded by $z^{(t'-\Gamma_n)^+}$, where $(\cdot)^+=\max\{\cdot, 0\}$. 
		
		Next, we upper bound the effect of truncating in each slot by introducing a modified version of the truncated strategy $\hat{\pi}_R^*$. Based on the relaxed scheduling strategy $\pi_R^*$, when $|\Omega(t)|>M$, the new truncated strategy $\hat{\pi}_R^*$ is designed by: instead of not scheduling a sensor because of limited bandwidth constraint, schedule it as $\pi_R^*$, but add a penalty $\sum_{t'=0}^\infty z^{(t'-\Gamma_n)^+}x_n(t)=(\Gamma_n+\frac{1}{1-z})x_n(t)$ on the total AoI. The AoI obtained by $\hat{\pi}_R^*$ will not decrease compared with $\tilde{\pi}$. Notice that when $|\Omega(t)|>M$, the $M$ sensors are chosen to be scheduled randomly. Thus, each sensor is not scheduled with probability $\frac{|\Omega(t)|-M}{|\Omega(t)|}$. Hence, the expected penalty add in slot $t$ can be upper bounded by:
		\begin{align}
			&\mathbbm{1}_{|\Omega(t)|>M}\sum_{n=1}^N(\Gamma_n+\frac{1}{1-z})x_n(t)\frac{|\Omega(t)|-M}{|\Omega(t)|}\nonumber\\
			\leq&\mathbbm{1}_{|\Omega(t)|>M}\sum_{n=1}^N(\Gamma_n+\frac{1}{1-z})x_n(t)\frac{|\Omega(t)|-M}{M}\nonumber\\
			=&\sum_{n=1}^N(\Gamma_n+\frac{1}{1-z})x_n(t)\frac{(|\Omega(t)|-M)^+}{M}
		\end{align}
		Let $x_n(t)$ be the AoI obtained by $\pi_R^*$ and $\mathbbm{1}_{(\cdot)}$ be the indicator function, then the difference between $J(\tilde{\pi})$ and $J(\pi_R^*)$ can be upper bounded as follows:
		\begin{align}
		&(J(\tilde{\pi})-J(\pi_R^*))\nonumber\\
		\leq&(J(\hat{\pi}_R^*)-J(\pi_R^*))\nonumber\\
		\leq&\sum_{n=1}^N(\Gamma_n+\frac{1}{1-z})x_n(t)\frac{(|\Omega(t)|-M)^+}{M}\\
		{\leq}&\frac{\max_{n}\Gamma_n+\frac{1}{1-z}}{MNT}\mathbb{E}\left[\sum_{t=1}^T\sum_{n=1}^Nx_n(t)\left(|\Omega(t)|-M\right)^+\right]\nonumber\\
		\overset{(a)}{\leq}&\frac{\max_{n}\Gamma_n+\frac{1}{1-z}}{MNT}\mathbb{E}\left[\sum_{t=1}^T\sum_{n=1}^Nx_n(t)\left(|\Omega(t)|-\overline{\Omega}\right)^+\right]\nonumber\\
		\overset{(b)}{\leq}&\frac{\max_{n}\Gamma_n+\frac{1}{1-z}}{\theta N^2T}\mathbb{E}\left[\sum_{t=1}^T\sum_{n=1}^Nx_n(t)\left||\Omega(t)|-\overline{\Omega}\right|\right]\nonumber\\
		\overset{(c)}{\leq}&\frac{\max_n\Gamma_n+\frac{1}{1-z}}{\theta N^2T}\mathbb{E}\left[\sum_{t=1}^T\sum_{n=1}^N\tau_{n, Q}||\Omega(t)|-\overline{\Omega}|\right]\nonumber\\
		=&\frac{(\max_n\Gamma_n+\frac{1}{1-z})\sum_{n=1}^N\tau_{n, Q}}{\theta N}\mathbb{E}\left[\frac{1}{NT}\sum_{t=1}^T||\Omega(t)|-\overline{\Omega}|\right],\label{eq:asymptotic1}
		\end{align}
		where inequality (a) is because inequality \eqref{OmegaUB} and (b) is because $(\cdot)^+\leq|\cdot|$. Inequality (c) is obtained because following the relaxed strategy $\pi_R^*$, each decoupled user has a set of activation thresholds, hence the AoI $x_n(t)$ cannot exceeds the largest thresholds in the worst channel $\tau_{n, Q}$.

		Finally, according to \cite{diaconis1991closed}, the expectation of $||\Omega(t)|-\overline{\Omega}|$ satisfies:
		\begin{equation}
		\mathbb{E}\left[\frac{1}{N}||\Omega(t)|-\overline{\Omega}|\right]=\mathcal{O}\left(\frac{1}{\sqrt{N}}\right).
		\label{eq:asymptotic2}
		\end{equation}
		
		Notice that the for users with fixed power constraint $\mathcal{E}_n$, the difference of threshold structure $\Gamma_n$ does not grow with the number of users in the network $N$. Moreover, $\frac{1}{1-z}=\frac{1}{\min_n\eta_{n, q}(1-\frac{N-M}{N})}=\frac{1}{\min_n\eta_{n, Q}\theta}$, which does not grow with the number of sensors $N$. In addition, $\frac{N}{M}=\theta$ suggests the available bandwidth $M$ will grow with the number of users $N$, thus $\tau_{n, Q}$ will not grow with $N$. Hence, the average of all thresholds:
		\begin{equation}
		\frac{1}{N}\sum_{n=1}^N\tau_{n, Q}=\mathcal{O}(1).
		\label{eq:asymptotic3}
		\end{equation}
		
		By combining inequalities Eq.~\eqref{eq:asymptotic1}-\eqref{eq:asymptotic3}, we will have the following upper bound:
		\begin{equation}
		J(\tilde{\pi})-J(\pi_R^*)=\mathcal{O}(\frac{1}{\sqrt{N}}).
		\end{equation}
		
		Considering that $J(\pi_R^*)$ is lower bounded by the performance of round robin policy $J(\pi^\text{RR})=\frac{1}{2}(\frac{N}{M}+1)$, which has no power consumption constraint. With $\frac{N}{M}=1/\theta$ is a constant and let $N\rightarrow\infty$, we can lower bound $J(\pi_R^*)$ by:
		\begin{equation}
		J(\pi_R^*)\geq J(\pi^{\text{RR}})=\frac{1}{2}(\frac{1}{\theta}+1).
		\end{equation}
		Finally, the asymptotic optimum performance of the proposed policy $\tilde{\pi}$ can be verified:
		\begin{equation}
		\frac{J(\tilde{\pi})-J(\pi_R^*)}{J(\pi_R^*)}=\mathcal{O}(\frac{1}{\sqrt{N}}).
		\end{equation}
	\end{IEEEproof}

\section*{Acknowledgement}
This work was supported in part by the National Key R\&D Program of China (Grant No.2017YFE0112300), Shenzhen basic Research Project (No.JCYJ20170816152246879) and the Tsinghua University Tutor Research Fund. The authors are grateful to Prof. Philippe Ciblat from Telecom Paris and Mr. Qining Zhang, Mr. Yuchao Chen, Mr. Jingzhou Sun from Tsinghua University for helpful suggestions. 

\end{document}